\pdfoutput=1 
\documentclass[aps,twocolumn,prl,showpacs,amsmath,amssymb,10pt]{revtex4-1}
\usepackage{graphicx}
\usepackage{scrextend}
\usepackage{manfnt,pifont}

\usepackage{url}

\usepackage{CJK}

\begin{document}


\begin{CJK*}{GB}{}
\CJKfamily{gbsn}

\title{All Tree-Level Correlators for M-theory on $AdS_7 \times S^4$}

\author{Luis F. Alday$^a$}
\author{Xinan Zhou (ÖÜÏ¡éª)$^b$}

\affiliation{$^a$Mathematical Institute, University of Oxford, Andrew Wiles Building, Radcliffe Observatory Quarter, Woodstock Road, Oxford, OX2 6GG, U.K.}
\affiliation{$^b$Princeton Center for Theoretical Science, Princeton University, Princeton, NJ 08544, USA}

\date{\today}

\begin{abstract}
\noindent 
We present a constructive derivation of all four-point tree-level holographic correlators for eleven dimensional supergravity on $AdS_7 \times S^4$. These correlators correspond to  four-point functions of arbitrary one-half BPS operators in the six-dimensional $(2,0)$ theory at large central charge. The crucial observation is that the polar part of the correlators in Mellin space is fully captured by a drastically simpler Maximally R-symmetry Violating (MRV) amplitude, while the contact part is fully fixed by superconformal Ward identities and the flat space limit. 
\end{abstract}

\maketitle
\end{CJK*}

\noindent{\bf Introduction.} Correlators of local operators in holographic CFTs are perhaps the most natural observables to test and exploit the AdS/CFT duality. In the regime where classical supergravity is a good approximation the computation is in principle straightforward. One needs to compute the effective action on $AdS_{d+1}$, obtained by the Kaluza-Klein reduction of the corresponding supergravity on $S^{D-d-1}$, and then 
sum over all the relevant Witten diagrams. In practice, however, this 
is a very cumbersome task. Recently, powerful methods have been developed that are 
inspired by 
the ones for flat space scattering amplitude. 
The analogy becomes manifest in Mellin space \cite{Mack:2009mi,Penedones:2010ue}, where these methods exploit efficiently the symmetries of the problem and the analytic properties of the Mellin amplitudes. 
This has led to an expression for all tree-level correlators in the case of $AdS_5 \times S^5$ \cite{Rastelli:2016nze,Rastelli:2017udc} and partial progress for $AdS_7 \times S^4$ \cite{Rastelli:2017ymc,Zhou:2017zaw}.  The latter background is particularly interesting, as the supergravity is the low energy limit of M-theory on $AdS_7 \times S^4$, and is dual to the 6d $(2,0)$ CFT at large central charge. 

The goal of this letter is to compute all tree-level correlators for M-theory on $AdS_7 \times S^4$ by borrowing another idea 
 from flat space. Consider the four-point correlator of the super primary of the stress tensor multiplet \cite{Arutyunov:2002ff}. In Mellin space it takes the form
\begin{equation}
\nonumber \mathcal{M}=\frac{P(s,t;\sigma,\tau)}{(s-4)(s-6)(t-4)(t-6)(u-4)(u-6)}
\end{equation}
where $s+t+u=16$ and $P(s,t;\sigma,\tau)$ is a complicated polynomial.
The key observation is that this amplitude simplifies drastically for a specific choice of the R-symmetry cross ratios
\begin{equation}
\nonumber \mathcal{M} \big|_{\sigma=0,\tau=1}=\frac{(u-8)(u-10)}{(s-4)}+\frac{(u-8)(u-10)}{4(s-6)}+(s\to t)
\end{equation}
We denote this amplitude as {\it Maximally R-symmetry Violating} (MRV), in analogy with MHV amplitudes in flat space. The presence of zeroes in the numerator follows from the fact that in this configuration low-twist long operators are not exchanged, and is a feature of all tree-level correlators. At the level of Witten diagrams the presence of these zeroes is non-trivial, and it organizes exchange diagrams into multiplets. This allows us to write the MRV correlator in terms of a set of already available cubic scalar vertices. 
We can further use R-symmetry to restore the $\sigma$, $\tau$ dependence and recover the full correlator up to the possible addition of contact terms, which can be 
fixed by superconformal Ward identities and the flat space limit. This procedure then leads to an expression for all tree-level correlators for M-theory on $AdS_7 \times S^4$.

\vspace{0.2cm}
\noindent{\bf Kinematics.} We focus on the one-half BPS local operators $\mathcal{O}_k^{I_1,\ldots,I_k}(x)$, $I_k=1\,,\ldots\,,5$ in the 6d $\mathcal{N}=(2,0)$ theory. These operators have protected conformal dimensions $\Delta_k=2k$, $k=2,3,\ldots$, and transform in the rank-$k$ symmetric-traceless representation of the $SO(5)$ R-symmetry group. We keep track of the R-symmetry structure by contracting the  indices with a null vector
\begin{equation}
\mathcal{O}_k(x,t)=\mathcal{O}_k^{I_1,\ldots,I_k}(x)t_{I_1} \ldots t_{I_k}\;,\quad t\cdot t=0\;,
\end{equation}
and denote the four-point correlation functions by
\begin{equation}
G(x_i,t_i)=\langle \mathcal{O}_{k_1}\mathcal{O}_{k_2}\mathcal{O}_{k_3}\mathcal{O}_{k_4}\rangle\;.
\end{equation}
Invariance under the conformal and R-symmetry group implies that  correlators depend only on the invariant cross ratios after a kinematic factor is extracted. Without loss of generality, we order the weights as $k_1\leq k_2\leq k_3\leq k_4$. It is necessary to distinguish two cases  
\begin{equation}
\nonumber k_1+k_4\geq k_2+k_3 \;\; \text{(case I)}\;,\quad k_1+k_4< k_2+k_3 \;\; \text{(case II)}\;.
\end{equation}
It is also useful to define {\it extremality} by
\begin{equation}
\mathcal{E}=\tfrac{k_1+k_2+k_3-k_4}{2} \; \text{(case I)}\;,\;\; \mathcal{E}=k_1 \;\; \text{(case II)}\;.
\end{equation}
The correlators can then be written as 
\begin{equation}
G(x_i,t_i)=\prod_{i<j}\left(\tfrac{t_{ij}}{x_{ij}^4}\right)^{\gamma^0_{ij}}\left(\tfrac{t_{12}t_{34}}{x_{12}^4x_{34}^4}\right)^{\mathcal{E}}\mathcal{G}(U,V;\sigma,\tau)
\end{equation}
where $x_{ij}=x_i-x_j$, $t_{ij}=t_i\cdot t_j$, and 
\begin{equation}
U=\tfrac{x_{12}^2x_{34}^2}{x_{13}^2x_{24}^2}\;,\;\;V=\tfrac{x_{14}^2x_{23}^2}{x_{13}^2x_{24}^2}\;,\;\; \sigma=\tfrac{t_{13}t_{24}}{t_{12}t_{34}}\;,\;\;\tau=\tfrac{t_{14}t_{23}}{t_{12}t_{34}}
\end{equation}
are the conformal and R-symmetry cross ratios. The exponents $\gamma_{ij}^0$ are given by
\begin{eqnarray}
&&\gamma_{12}^0=\gamma_{13}^0=0\;,\;\; \gamma_{34}^0=\frac{\kappa_s}{2}\;\; \gamma_{24}^0=\frac{\kappa_u}{2}\;,\\
\nonumber && \gamma_{14}^0=\frac{\kappa_t}{2}\,,\;\; \gamma_{23}^0=0\,,\;\text{(I)}\,,\;\;\gamma_{14}^0=0\,,\;\; \gamma_{23}^0=\frac{\kappa_t}{2}\,,\;\text{(II)}
\end{eqnarray}
where we have defined the shorthand notations
\begin{eqnarray}
\nonumber \kappa_s&\equiv&|k_3+k_4-k_1-k_2|\;,\;\; \kappa_t\equiv|k_1+k_4-k_2-k_3|\;, \\
\kappa_u&\equiv&|k_2+k_4-k_1-k_3|\;.
\end{eqnarray}
The correlator $\mathcal{G}(U,V;\sigma,\tau)$ is a polynomial of $\sigma$ and $\tau$ of degree $\mathcal{E}$. 

\vspace{0.2cm}
{\bf Construction.} The 6d $(2,0)$ theory arises as the worldvolume theory for a stack of $n$ M5-branes, where for large central charge $c \sim 4 n^3$. The four-point functions admit an expansion in $1/n$
\begin{equation}
\mathcal{G}=\mathcal{G}_{\rm disc}+n^{-3}\mathcal{G}_{\rm tree}+\ldots
\end{equation}
where the leading $\mathcal{G}_{\rm disc}$ is given by generalized free field theory. The sub-leading $\mathcal{G}_{\rm tree}$ can be computed from tree-level 11D supergravity on $AdS_7\times S^4$, as an expansion of exchange and contact Witten diagrams
\begin{equation}\label{GasGexchGcon}
\mathcal{G}_{\rm tree}= \mathcal{G}^{(s)}_{\rm exch}+\mathcal{G}^{(t)}_{\rm exch}+\mathcal{G}^{(u)}_{\rm exch}+\mathcal{G}_{\rm con}\;.
\end{equation}
The exchanged fields are organized into superconformal multiplets labelled by the Kaluza-Klein level $p$, where the global symmetry quantum numbers of the relevant component fields are summarized in the table below.
{\begin{center}\small
 \begin{tabular}{||c| c | c | c | c | c | c ||} 
 \hline
 field & $s_p$ & $A_{p,\mu}$ & $\varphi_{p,\mu\nu}$ &  $C_{p,\mu}$ & $t_p$ & $r_p$ \\ [0.5ex] 
 \hline\hline
$\ell$ & 0 & 1 & 2 & 1& 0 & 0\\ 
 \hline
$\Delta$ & $2p$ & $2p+1$ & $2p+2$ & $2p+3$ & $2p+4$ & $2p+2$ \\
 \hline
SO(5) &  $[p,0]$ & $[p-2,2]$ & $[p-2,0]$ & $[p-4,2]$& $[p-4,0]$ &  $[p-4,4]$\\ [0.5ex] 
 \hline
\end{tabular}
\end{center}}
\noindent We can therefore write 
\begin{eqnarray}\label{Vsp}
\mathcal{G}^{(s)}_{\rm exch}&=&\sum_p \mathcal{V}^{(s)}_p\;,\\
 \mathcal{V}^{(s)}_p&=&\lambda_s\, \mathcal{Y}_{[p,0]} W^{(s)}_{2p,0}+\lambda_A\, \mathcal{Y}_{[p-2,2]} W^{(s)}_{2p+1,1} \\
\nonumber&&+\lambda_{\varphi}\,\mathcal{Y}_{[p-2,0]} W^{(s)}_{2p+2,2}+\lambda_C\, \mathcal{Y}_{[p-4,2]} W^{(s)}_{2p+3,1}\\
\nonumber&&+ \lambda_t\, \mathcal{Y}_{[p-4,0]} W^{(s)}_{2p+4,0}+\lambda_r\, \mathcal{Y}_{[p-4,4]} W^{(s)}_{2p+2,0}
\end{eqnarray}
where $W^{(s)}_{\Delta,\ell}$ are the s-channel exchange Witten diagrams of dimension $\Delta$ and spin $\ell$. $\mathcal{Y}_{[d_1,d_2]}$ are R-symmetry polynomials associated with $SO(5)$ Dynkin labels $[d_1,d_2]$ of the exchanged fields, and their detailed expressions are given in the supplementary materials. $\lambda_{\rm field}$ are numerical coefficients which are related to the cubic vertices of the effective Lagrangian. R-symmetry selection rules and finiteness of the effective action require the summation range of $p$ to be finite, and even integer spaced
\begin{equation}\label{prange}
p-\max\{|k_1-k_2|,|k_3-k_4|\}=2\,,4\,,\ldots 2\mathcal{E}-2\;.
\end{equation}
Exchange contributions in other channels are similar, and are related to the s-channel by Bose symmetry. Finally, $\mathcal{G}_{\rm con}$ consists of contact diagrams with up to four derivatives, and all possible R-symmetry structures. Clearly, to follow exactly the standard diagrammatic expansion procedure is extremely cumbersome, if not impossible. One not only needs to find all the vertices - most of which are unknown \footnote{Cubic vertices for $s_k$ have been obtained in the literature \cite{Corrado:1999pi,Bastianelli:1999en}. Other vertices such as the general quartic vertices have not been worked out. By contrast, the general quartic vertices  in $AdS_5\times S^5$ were obtained in \cite{Arutyunov:1999fb}. But their complicated expressions took 15 pages.} - but also faces a proliferation of Witten diagrams.

In this letter, we point out a powerful organizing principle which exploits interesting properties of correlators in a special R-symmetry configuration. This allows us to constructively derive all four-point functions. To present our construction, it is most convenient to use the Mellin representation formalism \cite{Mack:2009mi,Penedones:2010ue}
\begin{equation}\label{inverseMellin}
\mathcal{G}_{\rm tree}=\int\limits_{-i\infty}^{i\infty} \frac{dsdt}{(4\pi i)^2} U^{\frac{s}{2}-a_s}V^{\frac{t}{2}-a_t}\mathcal{M}(s,t;\sigma,\tau)\,\Gamma_{\{k_i\}}
\end{equation}
where $a_s=k_1+k_2-2\mathcal{E}$, $a_t=\min\{k_1+k_4,k_2+k_3\}$, and 
\begin{eqnarray}
\nonumber \Gamma_{\{k_i\}}&&=\Gamma[\tfrac{2k_1+2k_2-s}{2}]\Gamma[\tfrac{2k_3+2k_4-s}{2}]\Gamma[\tfrac{2k_1+2k_4-t}{2}]\\
&&\times \Gamma[\tfrac{2k_2+2k_3-t}{2}]\Gamma[\tfrac{2k_1+2k_3-u}{2}]\Gamma[\tfrac{2k_2+2k_4-u}{2}]\,,
\end{eqnarray}
with $s+t+u=2\Sigma\equiv 2(\sum_{i=1}^4 k_i)$. In this language,  Witten diagrams have simple analytic structures. The exchange diagrams are a sum of simple poles 
\begin{equation}\label{Mellinexchange}
\mathcal{M}^{(s)}_{\Delta,\ell}(s,t)=\sum_{m} \frac{\mathcal{Q}_{m,\ell}(t,u)}{s-\Delta+\ell-2m}
\end{equation}
where $\mathcal{Q}_{m,\ell}(t,u)$ are degree-$\ell$ polynomials in $t$ and $u$. For $AdS_7\times S^4$, the above sum over $m$ always truncates, as a result of self-consistency at large $n$ \cite{Rastelli:2017udc}. Contact diagrams with $2L$ derivatives are polynomials in the Mandelstam variables of degree $L$. We should point out that the division into exchange and contact, such as in (\ref{GasGexchGcon}), is ambiguous. We can choose different on-shell equivalent cubic vertices in the spin-$\ell$ exchange, and the difference is only contact diagrams with up to $2(\ell-1)$ derivatives. 
In Mellin space (\ref{GasGexchGcon}) and (\ref{Vsp}) read
\begin{eqnarray}\label{Ssp}
&&\nonumber \mathcal{M}(s,t;\sigma,\tau)=\mathcal{M}^{(s)}_{\rm exch}+\mathcal{M}^{(t)}_{\rm exch}+\mathcal{M}^{(u)}_{\rm exch}+\mathcal{M}_{\rm con}\,,\\
\nonumber &&\mathcal{M}^{(s)}_{\rm exch}(s,t;\sigma,\tau)=\sum\nolimits_p \mathcal{S}^{(s)}_p(s,t;\sigma,\tau)\\
 &&\mathcal{S}^{(s)}_p=\lambda_s\, \mathcal{Y}_{[p,0]} \mathcal{M}^{(s)}_{2p,0}+\lambda_A\, \mathcal{Y}_{[p-2,2]} \mathcal{M}^{(s)}_{2p+1,1} \\
\nonumber&&\quad\quad+\lambda_{\varphi}\,\mathcal{Y}_{[p-2,0]} \mathcal{M}^{(s)}_{2p+2,2}+\lambda_C\, \mathcal{Y}_{[p-4,2]} \mathcal{M}^{(s)}_{2p+3,1}\\
\nonumber&&\quad\quad+ \lambda_t\, \mathcal{Y}_{[p-4,0]} \mathcal{M}^{(s)}_{2p+4,0}+\lambda_r\, \mathcal{Y}_{[p-4,4]} \mathcal{M}^{(s)}_{2p+2,0}\;.
\end{eqnarray}

Our key observation is that $\mathcal{M}(s,t;\sigma,\tau)$ simplifies drastically in a configuration where $t_1$ and $t_3$ are aligned. This corresponds to setting $\sigma=0$, $\tau=1$. In analogy with scattering in flat space, we term this limit Maximally R-symmetry Violating (MRV). The MRV amplitude 
\begin{equation}
\mathrm{MRV}(s,t)\equiv\mathcal{M}(s,t;0,1)
\end{equation}
has the following two distinguishing features
\begin{enumerate}
\item[{\it i})] there are only poles in the s- and t-channels;
\item[{\it ii})] there is a factor of zeros $(u-u_0)(u-u_0-2)$ with $u_0=2\max{\{k_1+k_3,k_2+k_4\}}$.
\end{enumerate}
The absence of u-channel poles is due to the R-symmetry suppression in the MRV configuration, where the u-channel R-symmetry polynomials vanish identically. The zeros manifest the decoupling of u-channel low-twist unprotected long multiplets in the MRV limit  \footnote{The first visible long operator exchanged has conformal twist $\mathfrak{t}=2\max\{k_1+k_3,k_2+k_4\}+4$, and corresponds to a double pole at $u=\mathfrak{t}$ in the integrand of (\ref{inverseMellin}). It is a super descendant of the double-trace super primary with smallest twist $\mathfrak{t}-4$, whose R-symmetry dependence is suppressed in MRV. Alternatively, these zeros can be seen from the difference operator $\hat{\Theta}$ in (\ref{reduced}) in the MRV limit. The reduced amplitude lacks poles to cancel such zeros in the full amplitude, as they would become unphysical away from the MRV limit.}. The zeros cancel one of the double poles from the $\Gamma_{\{k_i\}}$ factor, rendering logarithmic singularities in position space absent. Such singularities are tied to anomalous dimensions.

Remarkably, the zeros are satisfied individually by each super multiplet $\mathcal{S}^{(s)}_p$ (and similarly $\mathcal{S}^{(t)}_p$). This gives us the following simple strategy to compute correlators from Witten diagrams, which we outline below. Setting $\sigma=0$, $\tau=1$ and $t=2\Sigma-u-(\Delta-\ell+2m)$ in (\ref{Mellinexchange}) \footnote{This corresponds to a special choice of contact terms for exchange diagrams.}, the requirement of the u-channel zeros fixes {\it all} $\lambda_{\rm field}$ in (\ref{Ssp}) in terms of $\lambda_s$. This succinctly implements superconformal symmetry within each multiplet. Note that eliminating $t$ in terms of $u$ and the pole values of $s$, gives special Witten diagrams known as the ``Polyakov-Regge blocks'' \cite{Mazac:2019shk,Sleight:2019ive} (see also \cite{Gopakumar:2016wkt,Gopakumar:2016cpb,Gopakumar:2018xqi}). Such diagrams have improved u-channel Regge behavior ($s\to\infty$ keeping $u$ fixed). The coefficients $\lambda_s$ can be computed using the three-point coupling of $s_{k_1}s_{k_2}s_{k_3}$ obtained in \cite{Corrado:1999pi,Bastianelli:1999en} \footnote{Alternatively these three-point couplings  can be obtained by using the chiral algebra \cite{Beem:2014kka}. In fact,  the ratios of $\lambda_s$ can be fixed using superconformal Ward identities and the flat space limit, and only one cubic coupling is really needed. See footnote \cite{Note6}.}. The MRV amplitude is then simply the sum of s-channel and t-channel MRV amplitudes from each multiplet with no additional contact terms \footnote{Note that a contact term is at most linear in the Mandelstam variables, while the required u-channel zeros are degree 2. }
\begin{equation}
\mathrm{MRV}(s,t)=\sum\nolimits_p \mathcal{S}^{(s)}_p(s,t;0,1)+(\text{t-channel})\;.
\end{equation}
However, we learn much more than just the MRV limit. Since all R-symmetry polynomials in $\mathcal{S}^{(s)}_p$ are non-vanishing in the MRV limit, we can substitute the $\lambda_{\rm field}$ values into (\ref{Ssp}) and obtain all the singular part of the Mellin amplitude! It further turns out that there is a special choice of the polynomial residues such that {\it no} explicit contact terms are needed. This leads to a simple formula for the full Mellin amplitude, as we now present.

\vspace{0.2cm}
\noindent{\bf All four-point Mellin amplitudes.} The Mellin amplitude is the sum over three channels
\begin{equation}\label{MellinM}
\mathcal{M}=\sum_{i,j}\sigma^i\tau^j(\mathcal{M}^{i,j}_s(s,t)+\mathcal{M}^{i,j}_t(s,t)+\mathcal{M}^{i,j}_u(s,t))
\end{equation}
The three channel are related by Bose symmetry 
\begin{eqnarray}
\mathcal{M}^{i,j}_t(s,t)&=&\mathcal{M}^{i,\mathcal{E}-i-j}_s(t,s)\big|_{k_2\leftrightarrow k_4}\;,\\
\mathcal{M}^{i,j}_u(s,t)&=&\mathcal{M}^{\mathcal{E}-i-j,j}_s(u,t)\big|_{k_2\leftrightarrow k_3}\;,
\end{eqnarray}
and each channel is a sum over simple poles
\begin{equation}
\mathcal{M}^{i,j}_s(s,t)=\sum_{h=h_{\min}}^{h_{\max}} \frac{R^{i,j}_h(t,u)}{s-2 h}
\end{equation}
where $h_{\min}=\max\{|k_1-k_2|,|k_3-k_4|\}+2$, $h_{\max}=\min\{k_1+k_2,k_3+k_4\}-1$. The residues $R^{i,j}_h(t,u)$ are a sum over  supergravity multiplets labelled by the Kaluza-Klein level $p$ in the finite set (\ref{prange})
\begin{equation}
R^{i,j}_h(t,u)=\sum_p \mathcal{R}^{i,j}_{p,m}(t,u)\;, \quad p+m=h\;,\;\;  m\in\mathbb{N}\;.
\end{equation}
 As outlined above, we first use the MRV limit and scalar cubic couplings to determine $\lambda_{\rm field}$ from each multiplet. Then in $\mathcal{S}^{(s)}_p$ we will use the Regge-improved Polyakov blocks for the exchange amplitudes $\mathcal{M}^{(s)}_{\Delta,\ell}$, {\it i.e.}, with $t$ eliminated in terms of $u$ and $m$. Since we have at most spin-2 exchanges, the residues depend on $u$ quadratically in the form 
 \begin{equation}
u^2+\alpha(i,j;m,p)\, u+\beta(i,j;m,p)\;. 
\end{equation}
We can restore the Bose symmetry in $t$ and $u$ by eliminating $m$ from $\alpha$ and $\beta$ using $t$ and $u$. This prescription gives the following compact answer
\begin{equation}\label{residueR}
\mathcal{R}^{i,j}_{p,m}(t,u)= K^{i,j}_{p}(t,u)\, L^{i,j}_{p,m}\, N^{i,j}_{p}\;,
\end{equation}
where no sum over indices is intended.  
The first building block is a degree-2 polynomial in $t$ and $u$
\begin{eqnarray}
\nonumber K^{i,j}_{p}&=& 2i(2i+\kappa_u)t^-t^++2j(2j+\kappa_t)u^-u^+\\
\nonumber&+&2j(1-\kappa_u)t^+u^-+2i(1-\kappa_t)u^+t^-\\
\nonumber &+& \frac{1}{4}(2p-\kappa_t-\kappa_u)(2p-2+\kappa_t+\kappa_u)(u^- t^-+16ij)\\
\nonumber &+&(\kappa_u+\kappa_t-2p)(\kappa_u+\kappa_t+2p-2)(i t^-+ju^-)\\
 &+&8ij(t^+(\kappa_u-1)+u^+(\kappa_t-1))-8ijt^+u^+
\end{eqnarray}
where $u^\pm=u\pm \kappa_u-\Sigma$, $t^\pm=t\pm \kappa_t-\Sigma$. The second factor 
\begin{eqnarray}
 \nonumber L^{i,j}_{p,m}&=&\frac{\Gamma[\frac{k_1+k_2-p+1}{2}]\Gamma[\frac{k_3+k_4-p+1}{2}]\Gamma[\frac{k_1+k_2+p}{2}]}{\pi^{\frac{3}{2}}m!\, i!\, j!\prod_{a=1}^4\sqrt{(2k_a-2)!}\,\Gamma[2p+m]} \\
 &\times &\frac{(-1)^{i+j+\frac{1}{4}(2p-\kappa_t-\kappa_u)}\Gamma[\frac{k_3+k_4+p}{2}]}{ \Gamma[k_1+k_2-m-p]\Gamma[k_3+k_4-m-p]}
\end{eqnarray}
implements the truncation of poles $m_{\max}=\min\{k_1+k_2,k_3+k_4\}-p$ (also of KK levels $p$). Finally, the last factor reads
\begin{equation}
 N^{i,j}_{p}=\frac{2^\Sigma (2p-1)\Gamma[\frac{2(p-1)+\Sigma-\kappa_s-4l}{4}]}{2^6\Gamma[\frac{\kappa_u+2+2i}{2}]\Gamma[\frac{2(p+2)-\Sigma+\kappa_s+4l}{4}]\Gamma[\frac{\kappa_t+2+2j}{2}]}
\end{equation}
where $i+j+l=\mathcal{E}$. Remarkably, formula (\ref{MellinM}) with (\ref{residueR})  gives the {\it full} answer for the Mellin amplitudes. We have tested it against many examples obtained using the bootstrap method \cite{Rastelli:2017ymc,Zhou:2017zaw}, finding perfect agreement. The absence of extra regular terms can be understood in terms of superconformal symmetry and the flat space limit, as we will explain below. Let us mention that the sum over multiplets $p$ can also performed in closed form, and gives  hypergeometric series up to ${}_8F_7$. In the MRV limit, the residues can be expressed in terms a single very well-poised ${}_7F_6$ function. However, we leave the result as a sum to better manifest the analytic structure.

\vspace{0.2cm}
\noindent {\bf Flat space limit.} It is illuminating to study the Mellin amplitude in the flat space limit $s,t \to \infty$. In this limit we expect to recover the 11D graviton amplitude, for a particular choice of orthogonal kinematics. For a detailed discussion of the flat space limit in the 11D context, see \cite{Chester:2018aca,Chester:2018dga}. From our explicit results we find
\begin{equation}
\nonumber
\lim_{s,t\to \infty}{\cal M}(s,t;\sigma,\tau) = {\cal N}_{\{k_i\}} \frac{\Theta_4^{\text{flat}}(s,t;\sigma,\tau)}{s t u} P_{\{k_i\}}(\sigma,\tau)
\end{equation}   
where 
\begin{equation}
\Theta_4^{\text{flat}}(s,t;\sigma,\tau) = \left(t u + t s \sigma + s u \tau \right)^2
\end{equation}
with $s+t+u=0$ in the flat space limit. $P_{\{k_i\}}(\sigma,\tau)$ is a polynomial explicitly given by
\begin{equation}
\nonumber P_{\{k_i\}} = \sum_{\substack{i+j+k = \mathcal{E} -2 \\ 0 \leq i,j,k \leq \mathcal{E}-2}} \frac{(\mathcal{E}-2)!\, \sigma^i \tau^j}{i!\,j!\,k!\, (i+\tfrac{\kappa_u}{2})!\, (j+\tfrac{\kappa_t}{2})!\, (k+\tfrac{\kappa_s}{2})!}\;.
\end{equation} 
Note that up to a normalization factor, this is the same polynomial as would enter in the flat space limit in the $AdS_5 \times S^5$ case \cite{Rastelli:2016nze}. The normalization factor ${\cal N}_{\{k_i\}}$ can in principle be fixed from our result, but its form will not be important for us. The appearance of the factor $\Theta_4^{\text{flat}}(s,t;\sigma,\tau)$ can be understood in different ways. From the 11D perspective, it follows from the form of the supergravity tree-amplitude 
\begin{equation}
{\cal A}_{R,{\rm tree}}^{11}(p_i,\zeta_i) = \ell_{11}^9 \hat K \frac{2^6}{s t u}
\end{equation}
with $\hat K $ a universal kinematic factor, which depends on the 11D graviton polarization vectors $\zeta^\mu$ and momenta $p^\mu$. As explained in \cite{Chester:2018dga}, for 11D gravitons in the appropriate polarization $\hat K$ should contain the factor  $\Theta_4^{\text{flat}}$. From the 6D perspective, this factor was shown to follow from the flat space limit of the 6D  superconformal Ward identities. In \cite{Chester:2018dga} the case $k_i=k$ was analyzed. Here we see these results extend to arbitrary $k_i$.

\vspace{0.2cm}
\noindent {\bf Absence of contact terms.} Let us now explain the absence of extra contact terms in our final result. It turns out that in all cases the Mellin amplitude can be recovered from a `reduced' Mellin amplitude
\begin{equation}
\label{reduced}
{\cal M}(s,t;\sigma,\tau)  = \widehat \Theta \widetilde {\cal M}(s,t;\sigma,\tau)\,, 
\end{equation}
with $\widetilde {\cal M}(s,t;\sigma,\tau)$ a polynomial of degree $\mathcal{E}-2$ in $\sigma,\tau$, and $\widehat \Theta$ a complicated difference operator defined in \cite{Rastelli:2017ymc}. Although its explicit form will not be needed here, one can explicitly check that in the flat space limit $\widehat \Theta$ acts multiplicatively and reduces to 
\begin{equation}
\widehat \Theta~ \sim ~s t u\, \Theta_4^{\text{flat}}(s,t;\sigma,\tau)\;. 
\end{equation}
This in particular implies that in the flat space limit the reduced amplitude must take the form
\begin{equation}
\widetilde {\cal M}(s,t;\sigma,\tau)   = \frac{\widetilde P_{\{k_i\}}(\sigma,\tau)}{(s t u)^2} + \widetilde P_{sl}(s,t;\sigma,\tau) + \cdots
\end{equation}
where $\widetilde P_{\{k_i\}}(\sigma,\tau)= \widetilde{\cal N} P_{\{k_i\}}(\sigma,\tau)$ for an exactly computable overall factor $\widetilde{\cal N}$. The subleading correction $\widetilde P_{sl}(s,t;\sigma,\tau)$ is a polynomial in $\sigma,\tau$ of degree $\mathcal{E}-2$, and has degree -7 in $s,t$. Now here comes the interesting part. Plugging this expansion back in (\ref{reduced}) we obtain
\begin{equation}
\nonumber {\cal M}(s,t;\sigma,\tau) -  \widehat \Theta \frac{\widetilde P_{\{k_i\}}(\sigma,\tau)}{(s t u)^2}  = \widehat \Theta \widetilde P_{sl}(s,t;\sigma,\tau) + \cdots
\end{equation}
In particular, expanding the l.h.s. around the flat space limit we find that the leading term vanishes, and the subleading term should contain a factor $\Theta_4^{\text{flat}}(s,t;\sigma,\tau)$. We have checked that this is indeed the case for our explicit expression for ${\cal M}(s,t;\sigma,\tau)$, for many cases. The addition of any contact term, of the form $h(\sigma,\tau) s+ \text{crossed}$, would spoil this property. Had we chosen a different representation of the polar part of the Mellin amplitude, then this condition would fix precisely which contact term we need to add \footnote{The structure for the Mellin amplitude around the flat space limit described here could also have been used to fix the coefficients $\lambda_s$ that entered in our construction, up to an overall normalization.}.

\vspace{0.2cm}

\noindent {\bf Discussion.} In this letter, we obtained all  tree-level $AdS_7\times S^4$ four-point functions, by following a constructive procedure which starts from the zeros in the MRV limit. This procedure is, in a way, similar to the amplitude bootstrap from soft limits \cite{Cheung:2014dqa}. Our method can be applied to a variety of other backgrounds to construct, {\it e.g.}, all M-theory four-point functions in $AdS_4\times S^7$ \cite{Alday:2020dtb}, which was initiated in \cite{Zhou:2017zaw}.  Moreover, we can use the procedure to give a constructive proof the general result in $AdS_5\times S^5$ \cite{Rastelli:2016nze,Rastelli:2017udc}.  It would also be  interesting to explore similar simplifying limits of R-symmetry configurations for higher-point correlation functions \cite{Goncalves:2019znr}.

Our procedure did not make much use of the details of the reduced amplitude. In fact, when translating our results to the reduced amplitude, on a case by case basis, no extra structure appears to emerge. This is perhaps an indication that a slightly different definition of the reduced amplitude is needed. It would be interesting to study this in detail and see if hidden structures emerge, akin to what happens in related contexts \cite{Caron-Huot:2018kta,Rastelli:2019gtj,Alday:2019nin,Giusto:2020neo}. On a more practical level, (\ref{residueR}) contains a wealth of CFT data.  These data are useful for comparing with the numerical superconformal bootstrap \cite{Beem:2015aoa}, and are also essential for constructing more general one-loop correlators \cite{Alday:2020tgi}.

\vspace{0.5cm}
\noindent{\bf Acknowledgements.} We are grateful to S. Chester and L. Rastelli for comments on the manuscript. The work of LFA is supported by the European Research Council (ERC) under the European Union's Horizon 2020 research and innovation programme (grant agreement No 787185). The work of XZ is supported in part by the Simons Foundation Grant No. 488653.

\appendix

\section{R-symmetry polynomials}
In the following we give an expression for the R-symmetry polynomials $Y^{(a,b)}_{mn}(\sigma,\tau)$, following the setup in \cite{Nirschl:2004pa}. They are related to $\mathcal{Y}_{[d_1,d_2]}$ by $m=\frac{1}{2}(d_1+d_2-a-b)$, $n=\frac{1}{2}(d_1-a-b)$, and $a=\frac{\kappa_t}{2}$, $b=\frac{\kappa_u}{2}$. These are eigenfunctions of the $SO(d)$ Casimir operator.
\begin{equation}
\label{Casimir}
L^2 Y^{(a,b)}_{mn}(\sigma,\tau) = -2 C_{mn}^{(a+b)} Y^{(a,b)}_{mn}(\sigma,\tau) 
\end{equation}
where we define 
\begin{eqnarray}
\nonumber L^2 &=&2 {\cal D}_d^{(a,b)} -\tfrac{1}{2}(a+b)(a+b+d-2)\,,\\
C_{mn}^{(w)} &=& (n+\tfrac{w}{2})(n+\tfrac{w}{2}+d-3)+(m+\tfrac{w}{2})(m+\tfrac{w}{2}+1)\,, \nonumber
\end{eqnarray}
with ${\cal D}_d^{(a,b)}$ given in appendix B to \cite{Nirschl:2004pa}. We write the polynomials as an expansion in $\sigma$ and $\tau$
\begin{eqnarray}
Y^{(a,b)}_{mn}(\sigma,\tau) = \sum_{i,j=0}^{i+j=m} c_{i,j} \sigma^i \tau^j\,,
\end{eqnarray}
and recall $m \geq n$. The coefficients $c_{i,j}$ depend on $m,n,a,b$ and $d$, but we will leave this dependence implicit. Plugging this expansion into the Casimir equation (\ref{Casimir}) we obtain a recurrence relation for the coefficients $c_{i,j}$. For any fixed $m,n$ this recursion relation can be solved explicitly, but the problem becomes very complicated as $m,n$ increase. For this work we will be interested in getting analytic expressions for generic $m,n$ with $m-n$ finite. This can be done as follows. Solving a few cases one can note the following structure:
\begin{eqnarray}
\label{cansatz}
c_{i,j} = & &P^{(m-n)}(i,j)  \frac{(-1)^{m-i-j}}{ \Gamma (a+j+1) \Gamma (b+i+1)}    \\
& &\quad\quad \times \frac{ \Gamma \left(a+b+\frac{d}{2}+i+j+n-1\right)}{\Gamma (i+1) \Gamma (j+1) \Gamma (m-i-j+1)} \nonumber
\end{eqnarray}
where $P^{(m-n)}(i,j)$ is a polynomial in $i,j$ of total degree $m-n$. The overall normalization is fixed by the condition
\begin{eqnarray*}
P^{(m-n)}(m,0) =  \frac{\Gamma (a+1) \Gamma (m+1) \Gamma (m+b+1)}{\Gamma \left(a+b+\frac{d}{2}+m+n-1\right)}\;.
\end{eqnarray*}
This is equivalent to the condition that the highest power of $\sigma$ has coefficient $1$, namely $Y^{(a,b)}_{mn}(\sigma,\tau) = \sigma^m + \cdots$. Plugging (\ref{cansatz}) into the Casimir equation (\ref{Casimir}) we deduce a recursion relation for the polynomial $P^{(m-n)}(i,j)$
\begin{align*} 
&2 (i (2 a+b+d+4 j-2)+j (a+2 b+d-2)) P(i,j) \\
&-2 (m (a+b+d+m-3)+n (a+b+1)) P(i,j)  \\
& +2 (i^2+j^2-n^2)P(i,j)  =-2j(a+j) P(i+1,j-1) \\ 
&+(i+j-m) (2 a+2 b+d+2 (i+j+n-1)) P(i+1,j) \\
&+ (i + j - m) (2 a + 2 b + d + 2 (i + j + n-1)) P(i,j+1) \\
&- 2i(b+i) P(i-1,j+1)\;.
\end{align*}
For fixed $m-n$ this relation can be easily solved. If desired, the  R-symmetry polynomials $Y^{(a,b)}_{mn}(\sigma,\tau)$ for any fixed $m-n$ can be written in a closed form by using the $F_4$ Appell's generalised hypergeometric function
\begin{eqnarray*}
F_4\bigg[ \begin{matrix} a & b \\ c & d \end{matrix} ; x, y\bigg] =\sum_{m,n} \frac{(a)_{m+n} (b)_{m+n}}{m! n! (c)_m (d)_n} x^m y^n.
\end{eqnarray*}
Up to an overall normalization factor, they take the form
\begin{eqnarray*}
Y^{(a,b)}_{mn}= P(\sigma \partial_\sigma,\tau \partial_\tau) F_4\bigg[ \begin{matrix} -m & n+a+b+ \frac{d}{2}-1 \\ b+1 & a+1 \end{matrix} ; \sigma, \tau\bigg]
\end{eqnarray*}
with $P(i,j)$ the polynomial fixed by the procedure above.

\section{Exchange Mellin amplitudes}
We give the $AdS_7$ exchange Mellin amplitudes for spins up to 2 and generic conformal dimensions. These amplitudes can be obtained by solving the conformal Casimir equation, and are of the form
\begin{equation}
\mathcal{M}_{\Delta_E,\ell_E}(s,t)=\sum_{m}\frac{f_{m,\ell_E}\, Q_{m,\ell_E}(t,u)}{s-\Delta_E+\ell_E-2m}
\end{equation}
where $\Delta_E$ and $\ell_E$ are the dimension and spin of the exchanged field, and
\begin{eqnarray}
&&f_{m,\ell_E}=-\frac{2^{1-\ell_E}\Gamma[\Delta_E+\ell_E]}{m!(\Delta_E-2)_m \Gamma[\tfrac{\Delta^{1,2}_E+\ell_E}{2}]\Gamma[\tfrac{\Delta^{3,4}_E+\ell_E}{2}]}\\
\nonumber&&\quad \times \frac{\big(\tfrac{2-\ell_E-\Delta^{1,2}_E}{2}\big)_m\big(\tfrac{2-\ell_E-\Delta^{3,4}_E}{2}\big)_m}{\Gamma[\tfrac{\Delta^{1,E}_2+\ell_E}{2}]\Gamma[\tfrac{\Delta^{2,E}_1+\ell_E}{2}]\Gamma[\tfrac{\Delta^{3,E}_4+\ell_E}{2}]\Gamma[\tfrac{\Delta^{4,E}_3+\ell_E}{2}]}\;,
\end{eqnarray}
with $\Delta^{i,j}_k\equiv \Delta_i+\Delta_j-\Delta_k$. $Q_{m,\ell_E}(t,u)$ are polynomials in $t$ and $u$ of degree $\ell_E$, and are given by
\begin{eqnarray}
\nonumber Q_{m,0}&=&1\;,\\
\nonumber Q_{m,1}&=& \frac{(\delta_u^2-\delta_t^2)(t+u+4-\Sigma_\Delta)}{4(\Delta_E-5)}+(\Delta_E-1)(t-u)\;,\\
\nonumber Q_{m,2}&=& \frac{5T_1}{96(\Delta_E-6)}-\frac{5T_2}{96(\Delta_E-5)}+\frac{T_3}{96}\\
\nonumber &+&\frac{(\delta_u^2-\delta_t^2)}{2}(t-u)(u+t+4-\Sigma_\Delta)\\
\nonumber &-&\frac{2(1-\Delta_E+\Delta_E^2)-\delta_t^2-\delta_u^2}{12}(u+t+4-\Sigma_\Delta)^2\\
\nonumber &-&\Delta_E(1-\Delta_E)(t-u)^2\;,
\end{eqnarray}
where 
\begin{eqnarray}
\nonumber T_1&=&(\delta_u^2-\delta_t^2)(t+u+4-\Sigma_\Delta)\big(u(\delta_u^2-\delta_t^2-48)\;,\\
\nonumber &+&t (\delta_u^2-\delta_t^2+48)-(\delta_u^2-\delta_t^2)(\Sigma_\Delta-4)\big)\\
\nonumber T_2&=&((\delta_u-\delta_t)^2-4)((\delta_u+\delta_t)^2-4)(t+u+3-\Sigma_\Delta)\\
\nonumber &\times& (t+u+5-\Sigma_\Delta)\;,\\
\nonumber T_3&=&((\delta_u-\delta_t)^2-4)((\delta_u+\delta_t)^2-4)\\
\nonumber &+&8\Delta_E(\Delta_E-1)(2\Delta_E^2-12\Delta_E+38-\delta_t^2-\delta_u^2)\;,
\end{eqnarray}
and we have defined
\begin{eqnarray}
\nonumber\delta_t&\equiv&\Delta_1+\Delta_4-\Delta_2-\Delta_3\;,\\\nonumber\delta_u&\equiv&\Delta_2+\Delta_4-\Delta_1-\Delta_3\;,\\\nonumber\Sigma_\Delta&\equiv&\Delta_1+\Delta_2+\Delta_3+\Delta_4\;.
\end{eqnarray}

\vspace{-0.8cm}

\bibliography{refletter}
\bibliographystyle{utphys}

\end{document}